\documentclass[letterpaper]{article} 
\usepackage{aaai25}  
\usepackage{times}  
\usepackage{helvet}  
\usepackage{courier}  
\usepackage[hyphens]{url}  
\usepackage{graphicx} 
\usepackage{natbib}  
\usepackage{caption} 
\usepackage{algorithm}
\usepackage{algorithmic}

\usepackage{booktabs}
\usepackage{xcolor}

\usepackage{pdfpages}

\usepackage{newfloat}
\usepackage{listings}
\DeclareCaptionStyle{ruled}{labelfont=normalfont,labelsep=colon,strut=off} 
\floatstyle{ruled}
\newfloat{listing}{tb}{lst}{}
\floatname{listing}{Listing}
\nocopyright 

\title{Toward AI Matching Policies in Homeless Services:\\ A Qualitative Study with Policymakers}
\author{
    Caroline M. Johnston\textsuperscript{\rm 1}, Olga Koumoundouros\textsuperscript{\rm 1}, Angel Hsing-Chi Hwang\textsuperscript{\rm 1,\rm 2}, Laura
Onasch-Vera\textsuperscript{\rm 1}, Eric Rice\textsuperscript{\rm 1}, Phebe Vayanos\textsuperscript{\rm 1,\rm3,\rm 4}
}
\affiliations{
    \textsuperscript{\rm 1}CAIS Center for AI in Society, University of Southern California \\
    \textsuperscript{\rm 2}Annenberg School for Communication, University of Southern California\\
    \textsuperscript{\rm 3}Department of Industrial and Systems Engineering, University of Southern California\\
    \textsuperscript{\rm 4}Department of Computer Science, University of Southern California \\ carolinemjohnston39@gmail.com,
     \{koumound, hsingchi, onaschve, ericr, vayanou\}@usc.edu}

\begin{document}

\maketitle

\begin{abstract}
Artificial intelligence researchers have proposed various data-driven algorithms to improve the processes that match individuals experiencing homelessness to scarce housing resources. It remains unclear whether and how these algorithms are received or adopted by practitioners and what their corresponding consequences are. Through semi-structured interviews with 13 policymakers in homeless services in Los Angeles, we investigate whether such change-makers are open to the idea of integrating AI into the housing resource matching process, identifying where they see potential gains and drawbacks from such a system in issues of efficiency, fairness, and transparency. Our qualitative analysis indicates that, even when aware of various complicating factors, policymakers welcome the idea of an AI matching tool if thoughtfully designed and used in tandem with human decision-makers. Though there is no consensus as to the exact design of such an AI system, insights from policymakers raise open questions and design considerations that can be enlightening for future researchers and practitioners who aim to build responsible algorithmic systems to support decision-making in low-resource scenarios.
\end{abstract}

%

\section{Introduction}

As of January 2024, over 75,312 persons are experiencing unsheltered homelessness in Los Angeles (LA)~\citep{LAHSA_HomelessCount2024}. However, only 22,266 permanent housing units exist in the housing system, of which most are already occupied~\citep{LAHSA_HousingInventory2024}. Los Angeles uses the Vulnerability Index – Service Prioritization Decision Assistance Tool (VI-SPDAT), a 35-question self-reported survey about an individual's housing history; risks; socialization and daily functioning; and wellness, as part of prioritizing individuals for these scarce housing resources~\citep{Orgcode2014}. The VI-SPDAT measures an individual's vulnerability, where, in general, those who are more vulnerable (i.e., at higher risk of `adverse outcomes') are prioritized by human decision-makers who match individuals to resources and services. Though its use was well-intentioned, the VI-SPDAT has received great scrutiny in recent years due to evidence of racial bias within its administration and scoring results \cite{LAHSA_BlackPeople}. Additionally, the tool was never designed to match individuals to resources \cite{OrgCode_MovingForward}, resulting in misalignment between the system's goals and its use.

Concurrently, with the recent explosion of generative artificial intelligence (AI) tools such as ChatGPT,\footnote{https://chatgpt.com/} the general public is recognizing the power of AI in creating content, making predictions, and aiding in decision-making~\cite{Fried2024}. 
For example, in the homelessness field, AI researchers have created data-driven methods that prioritize or allocate scarce resources to individuals experiencing homelessness, potentially accounting for fairness (e.g., less racial bias) in the process \cite{Azizi2018, Toros2018, Kube2019, Rahmattalabi2022,Kube2023, Kumar2023, Tang2023}. 
However, the public has also become more aware of the harm caused by AI. Many negative perceptions of AI, including its ability to reinforce bias and its lack of transparency, persist within social services, creating a reluctance to adopt AI tools~\cite{Denton2019, AutomatingInequality, Reamer2023}. It remains to be seen whether AI tools that attempt to address potential fairness and transparency risks would be accepted within the homeless services community.

In this work, we aim to answer the following questions: \textit{Are homeless services policymakers in Los Angeles open to the use of AI-driven decision-support tools for matching individuals experiencing homelessness to housing resources? What are the perceived benefits and harms of pursuing this idea? What difficulties would lie ahead?} To this end, we conduct semi-structured qualitative interviews with 13 policymakers who directly influence the prioritization policies in the housing allocation system of LA to understand their perceptions of utilizing AI tools within the matching system.

Using the AI allocation policies of \citet{Tang2023} as a case study, we present potential efficiency, fairness, and transparency features of such a system and receive feedback on these topics. We focus on this particular AI system because of its use of causal inference techniques to optimally assign individuals to resources to maximize the number of individuals that successfully exit homelessness, subject to resource and fairness constraints. This aims to achieve a balance between efficiency and fairness in this highly constrained environment. Though it is not the only AI method in the homelessness field that uses causal inference~\cite{Kube2019, Rahmattalabi2022,Kube2023, Kumar2023}, it offers a transparent and responsive allocation policy, optimally assigning individuals whenever an individual or new resource enters the system, while others solve at a less granular and potentially delayed level, i.e., assigning resources on a weekly~\cite{Kube2019, Kube2023} or monthly~\cite{Kumar2023} basis.

As interest in research that informs AI governance grows, this work aims to guide how technologists can collaborate with policymakers to design systems that address real-world policy needs in the homeless services domain.
Specifically, we make the following contributions: 
\begin{enumerate}
    \item We reveal Los Angeles policymakers' openness to using AI tools to assist human decision-makers in the matching process and identify additional AI applications they view as beneficial for homeless services beyond matching.
    \item We identify potential gaps between what AI matching tools deliver and policymakers’ desired outcomes for fairness, efficiency, and transparency, based on a case study of a specific AI tool and interviews with policymakers in homeless services.
    \item We additionally identify potential implementation roadblocks for such systems.
\end{enumerate}  
The current study served as one of the very few that directly engaged with policymakers to extract actionable insights, complementing most prior work that studied perspectives of frontline workers and individuals experiencing homelessness~\cite{Kuo2023, Tang2024, Tracey2024a, Tracey2024, Moon2025}.
We also present recommendations and insights for AI researchers interested in bringing AI tools into these spaces.

\section{Literature Review}\label{sec:LitReview}
Various works involving algorithmic decision-making (i.e., AI) use qualitative methods to elicit feedback and insights on the design of such tools from relevant stakeholders. In public services, these studies range from child welfare \cite{Chouldechova2018, Saxena2021, Stapleton2022ImaginingStakeholders, Saxena2024} to job placement \cite{Moller2020,Flugge2021}, to criminal justice \cite{Brayne2021}, to refugee replacement~\cite{Ahani2021}. For example, \citet{Lee2019} develops a matching algorithm with a nonprofit that provides on-demand 
donation transportation through the use of volunteers, exploring the equity-efficiency tradeoff of such an algorithm.
Both \citet{Kuo2023} and \citet{Stapleton2022ImaginingStakeholders} use co-design activities to incorporate relevant stakeholders preferences and feedback into algorithmic processes in homeless services and child welfare, respectively. 

In recent years, AI researchers have studied how to integrate AI tools into the delivery of homeless services. \citet{Toros2018} develop a predictive algorithm to target and prioritize unhoused individuals who are predicted to have high future costs for public services. The authors argue that by prioritizing these high cost individuals for the most supportive services, then the incurred savings can be allocated to low cost individuals. Both \citet{Kube2019} and \citet{Tang2023} use causal inference techniques on observational data to design a resource allocation system for individuals experiencing homelessness. \citet{Kube2019} focuses on allocating resources to reduce the number of families who repeatedly enter homelessness while \citet{Tang2023} assigns clients to queues for resources to maximize the number of individuals that exit homelessness subject to potential fairness constraints. Taking the perspective of how to design prioritization policies subject to stakeholder preferences, \citet{Vayanos2020} designs a robust online preference elicitation algorithm and applies it to the setting of learning policymakers' preferences for potential trade-offs in efficiency and fairness metrics of housing allocation policies. 

Some have specifically studied algorithmic fairness with applications to homeless services.~\citet{Mashiat2022} examines fairness trade-offs in scarce resource allocation applied to administrative records from homeless services while~\citet{Akpinar2024} studies the fairness of public sector decision-making algorithms using under-reported administrative data.~\citet{Watson-Daniels2023} studies fairness inspired by scoring rules used to prioritize individuals experiencing homelessness for housing resources in Allegheny County (See~\citet{Showkat2023, Moon2024} for more complete surveys of AI in homeless services applications.)

The works listed above study methodological approaches of how to ``best'' or ``fairly'' use data to prioritize or allocate resources to individuals experiencing homelessness and do not interface with those using or affected by the proposed AI tools.  Thus, Human-Computer Interaction (HCI) researchers have begun to engage directly with the community in how its members interact with data and AI tools in homeless services \cite{Karusala2019, Kuo2023, Tang2024, Tracey2024a, Tracey2024, Moon2025}. For example, \citet{Kuo2023} engages with service providers and unhoused individuals about an already deployed data-driven prioritization tool that uses government administrative records instead of the typical, self-reported information from clients about their current housing situation. \citet{Tang2024} conducts workshops with service providers and unhoused individuals about ``AI Failure Cards,'' a way for the community to understand and provide feedback on the failures of a deployed AI housing allocation algorithm. 

Our work differs from these listed above in that we engage with policymakers, not frontline workers, in the discussion of the design of an AI allocation tool in the homeless services field. Because policymakers are more accustomed to ``systems level'' thinking, they are uniquely positioned to comment on topics such as the challenges of implementing AI across their organizations and the homeless services ecosystem. By using an AI resource allocation model case study to ground our discussion, policymakers are able to provide feedback that can be addressed or pursued head-on as opposed to post-deployment, if such a system were built.

\section{The Los Angeles Housing System}\label{sec:LAHousingSystem}
We now provide the necessary background of the LA housing resource allocation system.

\subsection{Vulnerability Assessment}\label{sec:VulnerabilityAssessment}
As a requirement from the United States Housing and Urban Development (HUD), the Los Angeles Housing Services Authority (LAHSA) has developed policy to prioritize and match individuals experiencing homelessness (clients) to available resources by their level of vulnerability~\cite{HUD2017}. 
To assess vulnerability, many housing service providers in LA will ask individuals to take the VI-SPDAT. From the client's answers to the VI-SPDAT, they receive points that gauge their risk
in a score of 0 to 17, where 0 is the lowest level of vulnerability and 17 is the highest level of vulnerability. In general, the survey is administered to a client at locations that help to connect them to housing services, such as shelters or through street outreach.

Historically, clients' VI-SPDAT scores were used to create an ordered priority queue 
to receive scarce housing resources. 
For example, a score between 8-17 would prioritize an individual by score for Permanent Supportive Housing (PSH) (e.g., affordable housing, case management services). 
In recent years, however, the community and researchers in homeless services have identified numerous issues in using the VI-SPDAT this way. This includes a racial bias in which White clients tended to score higher than non-White clients and findings that the VI-SPDAT does not accurately capture an individual's true level of vulnerability~\cite{LAHSA_matching}. Because of these complicating factors and a desire to prioritize individuals that are truly the most vulnerable, in more recent years, LAHSA's prioritization policy uses the VI-SPDAT score as an eligibility factor instead of an ordered priority and a revised version of the VI-SPDAT that more accurately assesses vulnerability~\cite{Rice2023CESTTRR}.

A client's VI-SPDAT answers and score, with other information such as their demographics and location, are recorded within the Home Management Information System (HMIS), an online database for collecting and accessing data on homelessness and housing resources. The HMIS database also records any resource that an individual is matched or connected with and attempts to track client outcomes. 

\subsection{Matching Clients to Housing Resources}\label{sec:Matching}
LA is divided into eight geographic regions called Service Planning Areas (SPAs). Each SPA is responsible for servicing its clients with its designated housing resources. Historically, each SPA had a small number of \textit{matchers} who matched clients to available housing resources on an individual basis, taking into account their VI-SPDAT score, information from their case worker, and eligibility requirements of the housing resource. For example, some PSH units are only available for veterans or those with an HIV/AIDS diagnosis, as set by the unit's particular funding source. 

As of July 2024, LAHSA has assumed a centralized responsibility for matching clients to PSH instead of the SPA matchers to streamline and increase visibility of the process~\cite{LAHSA_centralized_matching}.  
For the purposes of this work, considering an AI matching tool is still relevant and can most likely adapt to such a change.

\subsection{Known Critiques of the Current System}\label{sec:CritiquesCurrentSystem}
\paragraph{Efficiency} The VI-SPDAT was not intended for matching, something its creators have repeatedly noted, though various communities across the United States have adopted it as such~\cite{OrgCode_MovingForward}. Though LAHSA changed the way it is used within the matching process, as described in the previous section, 
the matching policy is not well-aligned in terms of improving potentially desired outcomes of the system. For example, if the goal of the policy is to get the most vulnerable individuals off the street, the VI-SPDAT is a flawed proxy to measure client vulnerability~\cite{LAHSA_matching} or risk of return to homeless services~\cite{Brown2018}, though there have been efforts to address its shortcomings~\cite{Rice2023CESTTRR}.

\paragraph{Fairness} Stakeholders at LAHSA and in homeless services at large are interested in promoting equity through homeless services. This is evident through LAHSA's changes to the administration and use of the VI-SPDAT in response to the evidence of racial bias in the tool.
In addition to these changes, to further investigate the challenges and obstacles of marginalized demographics of clients, LAHSA assembled an Ad Hoc Committee on Women and Homelessness and the Ad Hoc Committee on Black People Experiencing Homelessness~\cite{LAHSA_Women, LAHSA_BlackPeople}. 
The latter committee determined that though Black clients receive PSH at equal or higher rates compared to other racial groups, Black
clients return to homelessness at much higher rates. They also found that Black clients are disproportionately represented in the homeless population in LA.

\paragraph{Transparency} The housing system is a vast and complicated system (see \ref{sec:Results}{Results}), especially as policies, requirements, and responsibilities have shifted in recent years. 
As we will discuss, 
it is unclear whether clients accurately understand how the matching process works and the factors that facilitate their matching. Additionally, some of these processes and information are purposely withheld from clients to protect them, manage their expectations, or prevent them from ``gaming'' the system.

\section{Methodology}\label{sec:Methodology}
We now describe the methodology of our qualitative interviews and analysis. Our study draws inspiration from various works that engage with stakeholders to gain insight into the use and perceptions of AI tools in various domains such as homelessness \cite{Kuo2023, Tang2024}, food donation transportation \cite{Lee2019}, and child welfare \cite{Stapleton2022ImaginingStakeholders}.
Our interviews have policymakers elaborate on the shortcomings of the historical and current matching policies in regards to efficiency, fairness, and transparency. We then discuss a specific AI matching policy as a case study to inform discussions of how AI could be used for matching. Finally, policymakers express their attitudes and concerns about using AI in such applications.

\subsection{Recruitment \& Demographics of Interviewees} 
\begin{table*}
\centering
  \begin{tabular}[t]{lc}
    \toprule
    Demographic&Frequency (\%)  \\
    \midrule
    Advisory role representative & 77 \\ 
    Lead service provider representative & 23 \\ 
 \midrule
   Male & 23 \\
    Female & 69 \\ 
    Non-binary &  8\\ \midrule
     20-29 years of age & 8   \\
    30-39 years of age & 38  \\
    40-49 years of age & 31  \\
    50+ years of age & 23  \\ 
    \bottomrule
    \end{tabular}
     \begin{tabular}[t]{lc}
     \toprule
    Demographic&Frequency (\%)  \\
    \midrule
    1-4 years professional experience & 8   \\
    5-9 years professional experience & 23  \\
    10-14 years professional experience & 38  \\
    15+ years professional experience & 31  \\ \midrule
     White & 46 \\
    Black/African American & 23  \\ 
    Hispanic/Latino & 31 \\
    \midrule
    Experience as matcher or case worker  & 31\\ \midrule
    Lived homelessness experience & 15 \\
  \bottomrule
\end{tabular}
\caption{Aggregate Demographics of Participants ($n=13$)}
 \label{tab:Aggregate-demographics}
\end{table*}
We recruit 13 participants through a contact at LAHSA. These participants consist of policymakers in homeless services who either directly advise LAHSA or are representatives from leading service providers in particular SPAs. The group includes individuals with lived homelessness experience and employees of various homeless services providers, non-profits, and government agencies in LA, forming a diverse group of stakeholders in functions, experiences, and geography. These participants directly influence LAHSA's policies and procedures, including its prioritization policies for matching individuals experiencing homelessness to housing resources.

Though small in relative size, these policymakers come from \textit{a highly specialized and influential} group in homeless services who affect key policy decisions in LA. They have the political agency to either discourage or catalyze the adoption of AI matching tools 
at the system level. Thus, it is critical to gauge their openness and hesitations to such ideas.

To protect the confidentiality of participants, we present their aggregate demographic statistics in Table \ref{tab:Aggregate-demographics}.
If their affiliating organization allowed such a payment, participants received a \$100 Amazon gift card for their participation. Our Institutional Review Board (IRB) reviewed and approved the study's protocols and materials.

\subsection{Protocol} 
One of the researchers, who is an expert in AI and matching systems, conducted 60-75 minute semi-structured interviews with participants over Zoom. We first asked about their previous experiences with and exposure to any AI in either their personal or professional lives (e.g., ChatGPT, recommendation services such as Netflix, or voice-assistants). We then asked whether they had previously considered using AI on HMIS data to assist in making matching decisions. We additionally asked about the shortcomings of the current matching system regarding its efficiency, fairness, and transparency.  

\begin{figure*}[t]
  \centering
  \includegraphics[width=0.8\linewidth]{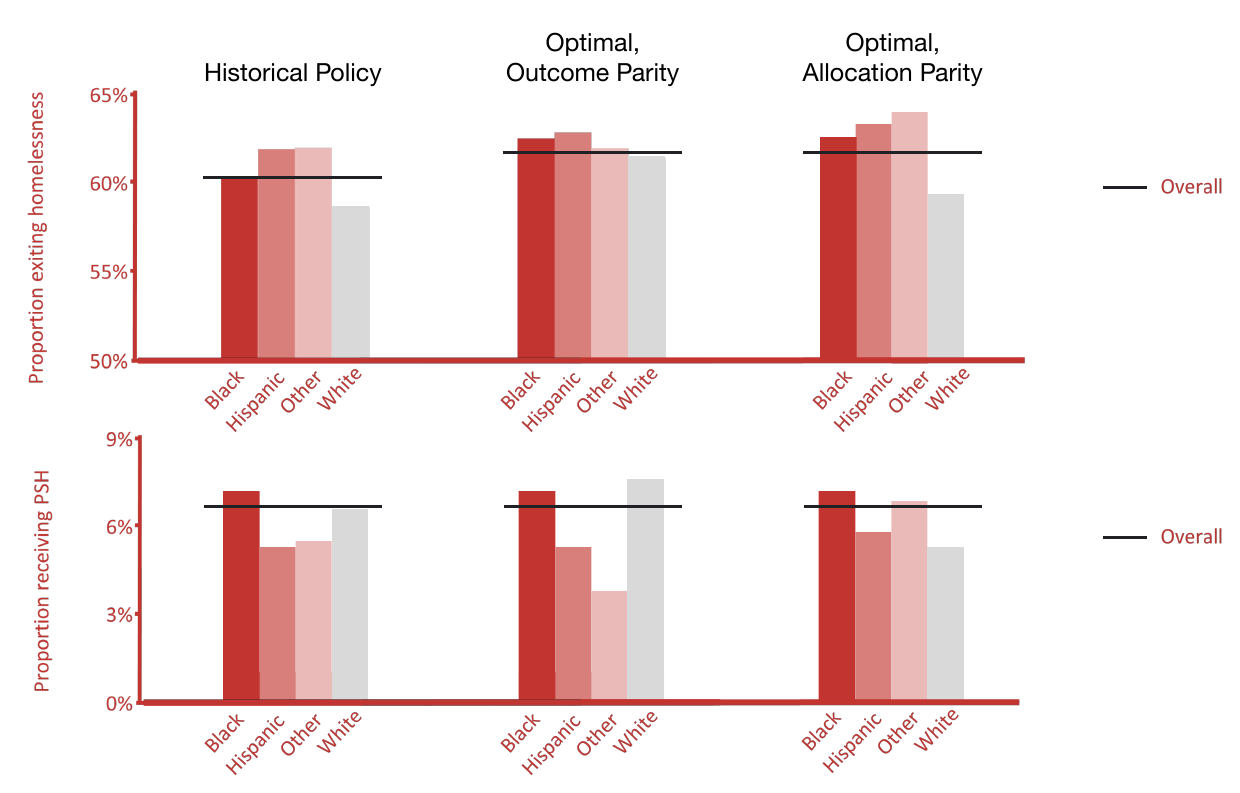}
  \caption{Fairness results of the historical allocation (``Historical Policy'') and the AI allocation policies trained to maximize the number of clients successfully exiting homelessness, showing the proportion of clients exiting homelessness and receiving PSH by race. The ``Optimal, Outcome Parity'' AI model ensures outcome parity across race while the ``Optimal, Allocation Parity'' AI model ensures allocation parity across race for PSH. The black line shows the proportion of clients exiting homelessness (resp. receiving PSH) over the entire client population. Both AI policies achieve a 2\% or 300 client increase in exits from homelessness. Adapted from~\citet{Tang2023}.} \label{fig:equity}
\end{figure*}

To discuss the potential benefits and drawbacks of using AI for matching more concretely, we presented the results and properties of the AI matching policies of~\citet{Tang2023}, trained on historical LA HMIS data, to the policymakers as a case study. The AI policies use causal inference techniques to assign clients to queues for a type of housing resource to maximize the expected number of individuals that successfully exit homelessness under the resource (or no resource) received (see the work itself for technical details). These queues include a ``services only'' queue (e.g., a day shelter or childcare services), a ``rapid rehousing'' queue (e.g., short term rental assistance), or a ``PSH'' queue. This assignment accounts for a limited budget or availability of each resource type and potential fairness constraints. 

We describe the transparency features of the AI policy to the policymakers in the following sense: the queues act in a ``first come, first served'' fashion where once a client enters the system (e.g., takes the VI-SPDAT), the AI policies use the client's HMIS information to assign them to a resource queue. When an available resource arrives in the system (e.g., a new PSH unit), it is presented to the first client in that respective resource line (e.g., the PSH queue). Clients that ``look the same'' in the data will be placed into the same queue, receiving the same type of resource or support. Because clients are in a queue, they or their case manager can know the client's place in line. Additionally, if the arrival rate of resources can be estimated (e.g., through expected construction timelines for PSH), clients or case managers can be provided with estimated wait times.

We additionally showed policymakers the potential gains in fairness (equity) by implementing these AI matching policies compared to the ``status quo,'' or historical matching policy. We focused on the following two metrics: 1) ``allocation parity,'' or ensuring that the rates that individuals receive housing resources are equal across protected groups and 2) ``outcome parity,'' or ensuring that the rates that individuals successfully exit homelessness are equal across protected groups. Particularly, we focus on race as the protected group for both metrics and the allocation rates of PSH as the resource in allocation parity. Previous work has shown that it is very unlikely for an allocation policy to achieve both allocation parity and outcome parity~\cite{JoTang_Allocation_Incompatability}, so we present results of two AI allocation policies to the policymakers, where one enforces allocation parity and the other enforces outcome parity  (see Figure \ref{fig:equity}).

We stress that our intent is not to see if the policymakers are interested in adopting the method of \citet{Tang2023}, but to use it as a concrete example of an AI matching policy to seek feedback on the properties of such a system. We did not focus on explaining the technical details of how the AI policies create the matching (i.e., causal inference and its assumptions), but instead focus more on what the AI policies could achieve in terms of it efficiency, equity, and transparency features. Finally, we asked for the policymakers' receptiveness and hesitations on adopting and implementing AI polices within the matching system in general. 
 Our detailed interview protocol can be found in the Appendix. 

\subsection{Qualitative Analysis}
The audio transcripts from the interviews amounted to 14 hours of qualitative data. To analyze these, we used a hybrid of inductive and deductive thematic analysis and used the qualitative data analysis software ATLAS.ti.\footnote{https://web.atlasti.com/} Two of the researchers, an expert in AI and an expert in Social Work, separately coded a subset of the transcripts using open coding. They then met to discuss and resolve differences in their interpretations of the data, considering their differing expertise and perspectives across disciplines, and formed a set of initial codes and codebook~\cite{Padgett2008}. They then coded another subset of transcripts and continued this process until they reached consensus after three total rounds of collaborative discussion and
coding. The AI researcher then independently coded the remaining transcripts using the agreed upon codebook.  Then the whole team met to give feedback on the codes and structure the themes for our thematic analysis. The finalized set of codes is listed in the Appendix. They include those that arose directly from the interview questions, e.g., ``Perceptions of AI," or how participants felt about integrating AI into the matching system, while others emerged organically from the discussions with the participants, e.g., ``Other AI Solutions," or areas beyond matching that policymakers envisioned that AI could assist with.

\subsection{Study Site}
We recognize that this study focuses on policymakers in LA and LA's specific housing resource matching process for individuals experiencing homelessness, which may not always generalize elsewhere. For example, not every community operates with a comparable SPA system. 
However, we believe this study contains insights that are applicable beyond LA. For example, as of 2024, many communities still use the VI-SPDAT in the process of prioritizing and/or matching clients including Santa Clara County, California~\cite{SantaClaraCES} and parts of Canada~\cite{Hamilton2024, CanadaHIFIS2024}. The issues discussed in our results 
such as messy data~\cite{Kube2023, Kuo2023, Moon2024, Moon2025}, bias from human decision-making~\cite{Chouldechova2018, Denton2019, Moon2024, Moon2025}, and high turnover rates in the workforce~\cite{Moses2023, KPMG2024} permeate across homeless services communities beyond LA. In fact, various other domains encounter these issues when implementing data-driven tools including policing and criminal courts~\cite{Brayne2021}; donation transportation services~\cite{Lee2019}; refugee resettlement~\cite{Ahani2021}; and job placement~\cite{Moller2020}.

 \begin{table*}[t]
  \begin{tabular}{cccccc}
    \toprule
Participant& \begin{tabular}[c]{@{}c@{}}Curr. use \\  AI?* (Q6) \end{tabular}& \begin{tabular}[c]{@{}c@{}}Sentiment of prev. \\ discussions of  AI* (Q6)\end{tabular}& \begin{tabular}[c]{@{}c@{}}Prev. considered \\ AI?\textsuperscript{\textdagger} (Q7)\end{tabular}&\begin{tabular}[c]{@{}c@{}} Sentiment of AI\textsuperscript{\textdagger} \\ pre-interview (Q8)\end{tabular}& \begin{tabular}[c]{@{}c@{}} Sentiment of AI\textsuperscript{\textdagger}\\  post-interview  (Q20)\end{tabular}\\
    \midrule
    P0 & Yes & Mix& Yes & Mix+ & Mix+   \\
    P1 & No& Mix- & No & Positive & Positive \\ 
    P2 & Yes & Mix& Yes & Mix+ & Mix+  \\
    P3 & Yes & Mix &No & Mix+ &Mix+\\  
    P4 & Yes&Mix & No & Mix & Positive\\  
    P5 &No & Mix & Yes & Negative & Negative\\  
    P6 &Yes &Mix & Yes &Mix &Mix\\ 
    P7 &Yes&Mix & No & Mix+ &Positive\\ 
    P8 &Yes & Mix &Yes & Mix &Mix+\\
    P9 & Yes& Mix & No &Mix&Mix\\  
    P10 & No& Positive & No & Mix & Positive\\   
    P11 & Yes& Mix & Yes & Mix & Mix+\\ 
    P12 & Yes& Negative & No &Positive &Positive\\ 
    
  \bottomrule
\end{tabular}
 \aboverulesep = 0.605mm
 \belowrulesep = 0.984mm
 \caption{Participants' previous experiences with and perception of AI ($n=13$). ``+'' (resp. ``-'') denotes that the participant indicated ``leaning'' towards a positive (resp. negative) sentiment when describing mixed feelings. See Appendix 
  for corresponding protocol question. *AI in general. \textsuperscript{\textdagger}AI for matching scarce resources to individuals experiencing homelessness. }  \label{tab:Policymaker-sentiments}
\end{table*} 

\section{Results}\label{sec:Results}
We stress that these participants are experts in homeless services, not in AI, and do not necessarily come from technical backgrounds. If participants had a question about the capabilities of AI, the AI expert interviewer would answer to the best of their ability. Misconceptions of AI by the participants were not necessarily corrected. 

We now present the specific findings from our analysis.  
In the following, we 1) assess the pre- and post-interview sentiments of participants concerning AI. We then 2) identify potential gaps between the AI matching model case study of \citet{Tang2023} and the real matching process as revealed by the policymakers. In other words, what remains to create a real AI matching system and are these requirements feasible? We then 3) discuss the limitations policymakers discussed with respect to the available data that may be used by AI.
 Finally, we 4) synthesize policymakers' openness and hesitations to using AI in homeless services both in general and with respect to the AI matching case study.

\subsection{Pre- and Post-Interview Results}
In Table~\ref{tab:Policymaker-sentiments}, we show participants' current level of exposure to and sentiments of AI, both in general and in terms of  matching individuals experiencing homelessness to scarce housing resources. Ten participants currently use AI in their daily life. 
Six, or less than half, of the participants had previously thought about how AI can be used to match individuals to scarce housing resources. Ten participants had mixed sentiments about using AI for matching before the interview. All participants either had the same or a more positive perception of AI after the interview, where four participants had a more positive perception. Only one participant had a negative view in both the pre-interview and post-interview.

\subsection{Reconciling an AI Matching Model with the Complex Reality of Matching}\label{sec:ReconcilingAIModel} 
The housing system in LA is complicated and political, as revealed below. Policymakers discussed logistical complexities within matching that are not reflected or accounted for within the AI matching case study. These include both external factors that are outside of the control of clients, case workers, and matchers; and internal complexities that must be addressed for an AI matching tool to create valid and effective matches. In fact, various policymakers envisioned a different or supplementary AI tool that can assist in some of these complex processes.

\subsubsection{Requirements for matching} 
Policymakers mentioned many factors when housing a client, ranging from eligibility for PSH units, to whether clients have a social security card. A client also needs responsiveness from their case worker to make the match, which ``\textit{is hard because points of contact change pretty frequently}'' (P6).
In terms of eligibility factors for resources, these can vary greatly due to the funder of the resource, ranging from age criteria for a senior complex or for Transition Age Youth (TAY), aged 18 to 24 years old; a mental health condition; or a requirement of chronic homelessness and  enrollment in a specific program such as Time-Limited Subsidies (TLS) (P10). In terms of incorporating these factors into an AI model to create valid matches,  P6 doubted that all of this data exists to facilitate such a tool:
\begin{quote}
    ``\textit{In theory, [the] data system has all of the different beds, the current data availability, the different eligibility criteria [...] all of that information that you need about that person. And so you can plug everything in, and it will pop out `This is the [housing] voucher that's available. This is the person.' But I don't think we have all of that information on the system side.}'' (P6)
\end{quote}

Multiple policymakers envisioned a supplementary AI tool to help caseworkers determine what resources their client is eligible for based on the resource's criteria and the characteristics of their client (P0, P6, P7, P11). P11 noted it would be helpful to use AI to find the ``\textit{most amount of PSH resources}" available that a client is eligible for. P0 described a system in which AI could notify case workers of missing, required forms. With this flag, case workers could kickstart the process of gathering the appropriate information. 

\subsubsection{No two resources nor clients are the same}\label{sec:NoTwoClientsSame} 
Many policymakers brought up idiosyncrasies of clients or resources that are not reflected in the AI matching case study of \citet{Tang2023}, which may be relevant for making matches. For example, P1 noted that in the HMIS, the exact disability and exact chronic health issues of clients are not necessarily recorded. However, these details may be relevant in regards to the level of services needed by that client:
\begin{quote}
    ``\textit{Maybe [a client's] disability is he doesn't have one leg, but he has a prosthetic, and his health issue might be asthma, [and he has] an inhaler ... [Another client's] disability could be schizophrenia and his chronic health issue might be COPD [Chronic Obstructive Pulmonary Disease], which is very different than asthma.}" (P1)
\end{quote}
Currently, though this information may not be in the HMIS, case workers may be aware and can make decisions accordingly.

Additionally, at an idiosyncratic level, not every client is at the same stage in their housing journey or needs the same type of support. Sometimes clients ``\textit{are not necessarily ready to be housed. Sometimes they need to go into interim housing ... [to] get a little structure in their lives. And then move on to living in a unit}'' (P5). P1 described how if a client has spent a long time in jail, then ``\textit{a congregate setting shelter is not appropriate. It's re-traumatizing because it reminds them of jail.}''  Another example is when housing a client who is a TAY: ``\textit{A 19 year old is not supposed to be housed in the same place from that time until they die. They move around}'' (P1). Thus, some housing options are not right for every client based on their experiences. Closely related to this is incorporating client preference or choice of resources, since clients ultimately need to accept or decline any resource offered (P0, P1, P6, P8). P8 mentioned that their organization has considered how an AI tool can match resources with respect to client preferences.

The case study treats each type of resource as functionally the same, though  P6 noted ``\textit{PSH is not equal across the system.}'' For example, older units may be in ``\textit{not as nice of an area that's not as well resourced}'' (P6). For TLS, some subsidies are only available for a certain period of time, while others can stay with the client for an extended period (P0). P8 described differences in resources in terms of the barriers clients may encounter: ``\textit{certain resources  themselves are harder to use between White people and people of color.}''

\subsubsection{External forces} 
Policymakers discussed various political factors that restrict the decision-making processes of the housing system. For example, the geographic boundaries of each SPA is determined by the LA County Department of Public Health. Within each SPA, there is a limited number of resources that their respective clients can access, where some SPAs do not build housing at all (P1). Thus, depending on a client's location, it can greatly limit their resource access. From a different political aspect, many policymakers mentioned that undocumented individuals cannot receive PSH, limiting their access as well (P3, P4, P6, P7, P11).

Politicians can also have direct influence over housing decisions, not just through legislation, but at the individual client level (P1, P3):
\begin{quote}
``\textit{If a politician calls you and says this has to happen, we have to do it. [...] It doesn't matter who [the client] and what their [VI-SPDAT] score is. We had to move forward with what they were wanting us to do.}'' (P1)
\end{quote}

\subsection{Data Issues: What Would AI Be Learning?}\label{sec:DataIssues}
The interviews with policymakers confirmed our understanding of the fidelity and lack of trust in relevant data. Policymakers voiced many concerns about the quality of housing data due to a myriad of issues, including human error and a lack of standardized data practices across or within agencies. This lack of trust threatens the efficacy of \textit{any} future AI system that would require training on historical data, whether it be used for matching or other applications.
\subsubsection{Messy and biased data}\label{sec:MessandBiasedDiata}
Many policymakers voiced a lack of trust in the recorded HMIS data (P0, P1, P2, P6, P8). Explicit descriptions of the data included ``\textit{horrible}'' (P0), ``\textit{messy}'' (P2), ``\textit{really crappy}'' (P6), ``\textit{not reflective of what we're seeing}'' (P7), and ``\textit{inaccurate a lot of times}'' (P11). Reasons for these characterizations included human error such as case managers not submitting or updating notes or not updating clients' information within the HMIS (P3, P4). 
\begin{quote}
{``\textit{Do we have total faith that all the data on an individual is accurate? If I were to build an AI model that would say who is the best match for it [PSH], you have to have some level of trust that whatever data gets fed into the AI learning model is good data. And I worry about that a little bit.}'' (P2)}
\end{quote}

Policymakers mentioned a lack of follow-up in client outcomes after allocating them a resource, details vital to building confidence in an AI matching system that utilizes machine learning assisted causal inference techniques (P6, P8): 
\begin{quote}``\textit{We'll connect someone somewhere and then we'll say `\emph{Check! Housed.}' But there's not a lot of data that says `\emph{Was this good?}' Even though we know that there are people who transition from one housing place to another housing place because they need to level up or down. I don't know that that is tracked at all.}'' (P6)
\end{quote}

There is worry that any AI system, whether used for housing decisions or other applications, would perpetuate biased data, as many policymakers directly referenced the evidence of racial bias in the VI-SPDAT 
(P1, P2, P5, P6, P9, P11):
\begin{quote}
    ``\textit{AI is only as smart as the data [...] so I think there's the potential for it [AI] to recreate a lot of challenges, [...] a lot of potentially ingrained racist policies of the past.}'' (P6)
\end{quote}
    Beyond the VI-SPDAT, P0 mentioned existing racial biases in who receives services through street outreach programs, where Hispanic individuals receive services at lower rates than they are experiencing homelessness: ``\textit{A lot of times we want to focus on the matching and the housing aspect of it, but that is only informed by who housing navigation is actually delivering services to.}'' Thus, there exists a racial bias in who is (or is not) even represented within the data.

Policymakers noted bias due to
human elements of housing. For example, newer case managers may lack training in recording data or administering the VI-SPDAT (P1, P3, P8). 
Or perhaps a client ``\textit{gets lucky}'' with a better case worker (P2, P11). Various policymakers noted a bias since data (e.g., the VI-SPAT, case notes, etc.) is collected through a ``\textit{human-interaction}'' process and can be subjective or misinterpreted (P2, P3, P8, P11). There is also doubt that the recorded VI-SPDAT answers are accurate. Clients may know that a higher score leads to housing, and case managers want to get their clients housed, leading to potentially ``\textit{fluffed up}'' answers (P4) and inflated VI-SPDAT scores (P4, P6, P12).  

\subsubsection{Lack of data practices}\label{sec:LackDataPractices}
Policymakers mentioned a culture that does not require or encourage standardized data practices (P1, P4). In particular, P1 noted that this is highly variable from agency to agency. Thus, if an AI matching system were introduced, this could lead to even more bias due to disparities in data quality:
\begin{quote}
    ``\textit{Some of the agencies that don't look at HMIS as a positive thing, their participants would lose out. Because we have small agencies, big agencies, grassroots [...] and it would have to be this universal agreement that this [data quality] is the priority, because we're going to be getting these [prioritization] lists [...] And so if you didn't do it right, your people might miss out.}'' (P1)
\end{quote}

P1 noted the lack of attention to data may be a result of social work culture, as ``\textit{social workers aren't great at math [...] so when you're asking `\emph{people people}' that only want to interface with people, data falls to the bottom.}''  However, they noted that their particular agency has had a stronger push toward data practices within the past few years.

\subsubsection{AI Tools to Improve Data Processes}
Beyond AI for matching, policymakers envisioned various other AI applications that could assist in the data challenges (P6, P7, P8, P12). For example, P7 mentioned that their organization is
using AI to ``\textit{make data more accessible for providers.}'' P12, in discussing that answers to the VI-SPDAT may not always be truthful,
suggested a way of using AI to populate certain information based on verifiable data. 

\begin{quote}
     ``\textit{I honestly don't have full understanding of how LAHSA manages their data right now, but it doesn't feel nimble [which] makes it difficult to make decisions and to gauge whether equitable outcomes exist [...] So any way that AI could help, even if not with the decision-making part of it, but the way that we are able to work with data [...] I think that that could be really huge.}'' (P8)
\end{quote}

\subsection{Perceptions of AI: the Good, the Bad, and the Unsure}\label{sec:PerceptionsofAI}
In general, with the discussion of the case study of \citet{Tang2023}, policymakers were excited about the possibilities of incorporating AI tools in the matching process to improve efficiency and remove human bias. They also voiced concern and doubt in how AI could address or be integrated into various aspects of the housing system. As to the specifics of the case study, policymakers valued the proposed equity gains but did not arrive at a clear consensus as to the merits and practicality of the proposed transparency aspects. We first discuss policymakers' thoughts related to incorporating any AI tool into the matching system and then their thoughts on the specifics of the \citet{Tang2023} case study.

\paragraph{Benefits of Using AI for Matching}

Policymakers viewed AI as a tool that could reduce the amount of tedious work often required of overburdened caseworkers with high turnover rates (P0, P3, P4, P6, P9, P10, P11): 
\begin{quote}
``\textit{I honestly think that [AI] can maybe help improve the epidemic that we're having right now [...] It can help [matchers] be less stressed out and actually value what they're doing [...] From what I see is they're overworked or overtired, and to have a tool that can help them do what they need to do would be very beneficial.}'' (P9)
\end{quote}
P12 noted the potential benefit of efficiency, where ``\textit{[with AI] it takes [fewer] people to figure out things, which is maybe replacing jobs. But maybe [these are] jobs that are hard to fill and people aren't staying in [...] It could potentially help the system move things faster.}'' In general, many policymakers thought AI could help those in homeless services be more ``efficient'' or to make parts of the housing process ``simpler'' (P0, P2, P3, P6, P7, P8, P10, P11, P12). 

Policymakers voiced that they were hopeful AI could reduce some human bias in the matching process.
Though they recognized that humans add bias in various ways, they still want humans involved, likening AI to an extra resource to consider as part of the matching process (P0, P1, P2, P3, P4, P6, P7, P9, P11, P12). As P11 noted, ``\textit{because  the work we're doing is human service work [...] I would hate to see it completely  gone, the human aspects to the mix.}''

\paragraph{Risks of using AI for matching}
Policymakers had various doubts that AI could replace all parts of the housing process, even though this is not necessarily the intent. For example, P1 noted that sometimes a particular SPA will not have enough PSH units: ``\textit{You have to go and beg other SPAs, `\emph{Can we please get some units?}' And that again is one of those things that AI can't [do].}'' P5 strongly opposed the use of AI for matching clients, noting ``\textit{I just can't see a computer doing something that is supposed to be a human connection. And it's gonna sever that.}''

There was a worry of becoming reliant on any AI system, noting that a tool is only effective if it is actually embraced by the end users. Thus, it is important to consider the larger decision-making environment
(P6, P9):
\begin{quote}
    ``\textit{People tend to follow them [guidelines] because they think they don't know all the history. And so, you might find that it's a suggestion or a preference that becomes a rule [...] So I think that's one of the dangers ...
    [Case managers are] making these judgments in a very short window of time with a lot of stress. So it's also the culture of who's utilizing the output that can also complicate.}'' (P6)
\end{quote}

\paragraph{Divergent views on equity}
In general, policymakers appreciated the potential equity gains under the system of \citet{Tang2023} as shown in Figure~\ref{fig:equity} (P2, P3, P4, P10, P12).
They did not agree as to whether the AI system should focus on outcome parity or allocation parity as the preferred metric. For example, P0 noted that historical biases in the data could skew the outcome parity model: ``\textit{What if historically we've just given resources to White and Black folks, and not Latin folks or Asian folks? So then they get the short end of the stick because ... [AI is] building off of whatever happened historically.}'' Instead, they ``\textit{felt better}'' about the allocation parity model (P0). Meanwhile, when discussing the allocation parity model, P7 stated ``\textit{not every person needs permanent supportive housing [...] I think that when we create AI, [AI developers] have to be able to understand [that]}.''

\paragraph{Challenges of operationalizing transparency} 
Policymakers did not agree, even within themselves, as to whether they saw the transparency aspects of the AI matching system of \citet{Tang2023} as an improvement over the current system. Even if they did appreciate such aspects, they grappled with how the proposed system might operationalize in practice (P1, P2, P3, P6, P8, P11, P12). For example, P1 noted  ``\textit{I would love it if we could say, `\emph{Look, you're this place in line, and it's estimated this time.}' I just don't understand how we can do that in our system, the way it's built right now.}'' 
P12 shared similar mixed sentiments, but recognized the needs of clients versus their own sentiments:
\begin{quote}
    ``\textit{I'm not certain how I feel about it [clients knowing their place in line].  However, if I were homeless, then I would want to know when might I have a housing resource? I can see why people would want to know that. But I also think it's somewhat dangerous because there are so many factors [...] The system is very complex.}'' (P12).
\end{quote}

Stakeholders compared the proposed queue model for housing resources to settings that utilize physical queues such as a ``\textit{deli}'' (P2) or a ``\textit{supermarket}'' (P6). They noted however, these models may not translate well to the housing setting. For example, P2 stated that one can get an estimated wait time for an order at the deli but in housing ``\textit{you're talking about something that's much higher stakes than a sandwich}.'' P6 additionally pointed out that it is easier to implement queues where there is a physical line instead of in the housing system where clients are in line ``\textit{on paper.}''

Policymakers noted concerns in providing particular information, such as estimated wait times, to clients directly, as opposed to their case worker. For example, if the housing timeline does not pan out, for whatever reason, this leads to disappointing clients and ``\textit{breaking their trust}'' (P0).  However, we note that sharing this information with clients is contentious within the current housing system (P0, P1, P2, P6, P10). 
As P6 explained, clients will ``\textit{be less likely to want to engage with the government if [they] knew that it would take [them] 10 years to get anything.}''
Similarly, P10 voiced that if estimated wait times are shared with clients, they may become less likely to fully engage in their own housing stabilization efforts. For example, they may not feel compelled to continue meeting with their mental health worker since they are ``\textit{gonna be next [for housing] really soon}'' (P10).

\section{Discussion}\label{sec:Discussion}
Our study engaged with policymakers in homeless services in LA, discussing the current challenges of the matching process and their openness and hesitations to using AI to assist in this process. Using the AI matching tool of~\citet{Tang2023} as a case study, we presented potential efficiency, equity, and transparency features that could be achieved through an AI matching system, gaining feedback on any value policymakers saw in these. We also discussed potential worries about using AI within the matching process. 
We now discuss the implications of these findings.

\subsection{Adopting AI tools in homeless services}
Our findings show that even with the awareness of data issues and potential risks to using AI, policymakers are willing to engage with and explore AI tools that are thoughtfully designed and implemented to improve the matching process for both service providers (matchers and case workers) and clients. As seen in Table \ref{tab:Policymaker-sentiments}, by the end of the interview, all but one policymaker recognized that AI could provide some positive change in the matching process. This sentiment holds even if all the implementation details are not finalized, as seen within our case study, demonstrating an openness to explore the idea in general. 

Though the interview questions were mainly designed to discuss AI within the matching process and the case study of \citet{Tang2023}, policymakers offered insights into different applications within or beyond matching where they envisioned AI assisting in homeless services. This provides multiple avenues of exploration for collaboration between AI researchers, developers, and those in homeless services to create such tools and assistants.

However, to build an AI matching tool that is readily adopted by the homeless services community,  AI researchers must address multiple issues. 
We believe these issues are relevant to public services in general.

\paragraph{Improving data quality and practices} 

There is currently an effort in LA to both utilize an updated version of the VI-SPDAT that more accurately assesses vulnerability and leverage additional, relevant data sets (i.e., county health data)~\cite{Rice2023CESTTRR}. Previous AI research has also shown that messy or missing data 
can potentially be corrected or accounted for, see e.g., \citet{Justin2023, Bertsimas2024_DataImputation, Kern2024_MissingDataCATE}.

Beyond these potential avenues, 
there are no standardized data practices across service providers in LA. With many competing day-to-day priorities and a lack of data culture as mentioned above, it may be difficult for those in public services to understand and justify dedicating resources to this effort. By creating proof-of-concept AI tools, AI researchers can show the ``value'' of prioritizing data in public services to policymakers who have the ability to affect such changes.

To create more faithful data for an AI matching system and to reduce potential biases across providers with different data standards, the adoption of uniform data practices is imperative. AI researchers must collaborate with service providers to define and integrate data quality standards and practices, potentially changing the culture within providers to prioritize these habits, as was previously discussed. For example, case workers can be trained on how to most accurately input client data and ``best practices'' if doing so is not feasible. In particular, unobserved confounders are of concern for causal inference methods such as~\citet{Tang2023}, so it is important that all information used in the matching process is recorded. One potential avenue for exploration is working closely with agencies that have been more successful at integrating data practices into their workflow. 

\paragraph{Building trust through evaluation} Even when thoughtfully designed, AI systems can perform differently when deployed due to issues such as shifting distributions of data~\cite{Justin2023, Kern2024_MissingDataCATE}, unobserved confounders~\cite{Kallus2021}, and changing system requirements as seen with LAHSA.
To build community trust in an AI matching system, AI researchers must periodically audit its performance to measure if it performs as desired by service providers and calibrate its performance as needed. From a data perspective, this requires a commitment to long-term follow up with clients (e.g., more than 2 years after receiving or not receiving a resource) to measure whether clients are staying housed with their given resource. 

From an operational perspective, evaluation should also involve follow-up (e.g., interviews) with how matchers engage with such an AI tool. This includes determining how often they follow the suggested matches compared to their on-the-ground expertise. Ideally, when matchers make decisions that differ from the AI system, this information can be used to further refine or train the model, moving towards a more useful tool for matchers. 

\paragraph{Multi-stakeholder engagement and agreement} 
Policymakers did not agree on what the AI system of \citet{Tang2023} should achieve. For example, there was no consensus in whether the AI matching policy should achieve allocation or outcome parity, or whether there are merits to using a first come, first served queue model. AI researchers must continue to engage with the community to determine the ultimate, desired properties of an AI matching system, even beyond the properties presented here. Though there is rich research in this area across public services, particular communities may desire different algorithmic design principles, thus, warranting this investigation \cite{Lee2019, Johnston2020, Stapleton2022ImaginingStakeholders, Johnston2023, Kuo2023}. This could involve additional input from current or former clients in addition to those who are policymakers and service providers.

\subsection{Limitations and Future Work}
Policymakers overwhelmingly voiced that if AI is integrated into the matching process, a human decision-maker should still be involved. Thus, a human matcher can use an AI matching tool as a suggestion or can ``override'' the tool. However, the results of Figure~\ref{fig:equity} are shown in regards to a decision-maker who follows the AI recommendations faithfully. Thus, it remains to be seen what efficiency, equity, and transparency gains are achieved with the integration of human decision-making and AI and how AI can further learn from the human decision-making process. 

Another limitation of this work is that a minority of interviewees have lived homelessness experience (see Table \ref{tab:Aggregate-demographics}). 
It is important to continue to engage with and educate individuals in this community on the potential applications, merits, and downsides of using AI within the matching process and homeless services in general since this is the population whose livelihoods would be most affected by such a tool. This invites the study and exploration of new methods to engage with the community such as those proposed in \citet{Kuo2023}. A related area to explore is that certain subpopulations of clients, such as victims of domestic violence, need additional consideration around issues of privacy and confidentiality if using their personal data in an AI system.

As noted previously, though our study participants are experts in homeless services in LA, issues of data quality, human bias, and high staff turnover are not unique to LA or even homeless services. Though relevant stakeholders' willingness to accept data-driven tools must be established in other locations and domains, continued research efforts in building data-driven decision assistance tools that address these challenges are an important step in creating tools that may benefit these communities.

\section{Conclusion}
    
In this work we conducted semi-structured qualitative interviews with 13 policymakers in homeless services in LA on current issues with the housing resource allocation system and their perceptions of utilizing AI tools in this space. Using the AI allocation policy of \citet{Tang2023} as a case study, we presented potential efficiency, fairness (equity), and transparency features of such a system and received feedback on these topics. We found that policymakers are in general open to the use of AI tools that are thoughtfully designed to assist in the matching process if human-decision makers are still involved. We presented various avenues for AI researchers to explore to engage with experts in homeless services in bringing AI tools into this domain.

\section{Positionality}
Our research team is an interdisciplinary mix of academics in Social Work, HCI, AI, and Operations Research (OR). Most of the team has previous research experience and community engagement experience with individuals in homeless services or those with lived homelessness experience. The lead researcher is an OR and AI specialist who conducted  the semi-structured interviews and participated in the coding process. Additionally, one of the social work researchers with experience with individuals experiencing homelessness participated in the coding process. The principal investigator is an expert in OR and AI with expertise in advancing AI/OR in directions that advance social justice and support the most vulnerable and under-served communities. They have been collaborating with social work researchers with expertise in homelessness and policymakers in homeless services for a decade. Three of the researchers identify as White women, one identifies as a White and Hispanic/Latino woman, one identifies as an Asian woman, and one identifies as a White man.

\section{Acknowledgments}
The authors gratefully acknowledge Hailey Winetrobe Nadel for her assistance and the Social Work Ph.D. students of the USC Center for AI in Society for their feedback on the interview protocol. They would also like to thank Marina Flores, Bryan Paz, and Paola Casillas for their help in subject recruitment and the participants of the study for their time and insight. P. Vayanos is funded in part by the National Science Foundation under CAREER award number 2046230. C.M. Johnston is funded through the Boucher Charitable Gift Fund.

\bibliography{references}

\begin{thebibliography}{56}
\providecommand{\natexlab}[1]{#1}

\bibitem[{Ahani et~al.(2021)Ahani, Andersson, Martinello, Teytelboym, and Trapp}]{Ahani2021}
Ahani, N.; Andersson, T.; Martinello, A.; Teytelboym, A.; and Trapp, A.~C. 2021.
\newblock Placement Optimization in Refugee Resettlement.
\newblock \emph{Operations Research}, 69(5): 1468--1486.

\bibitem[{Akpinar, Lipton, and Chouldechova(2024)}]{Akpinar2024}
Akpinar, N.-J.; Lipton, Z.~C.; and Chouldechova, A. 2024.
\newblock {The Impact of Differential Feature Under-reporting on Algorithmic Fairness}.
\newblock In \emph{Proceedings of the 2024 ACM Conference on Fairness, Accountability, and Transparency}, 1355--1382.

\bibitem[{Azizi et~al.(2018)Azizi, Vayanos, Eric, and Milind}]{Azizi2018}
Azizi, M.~J.; Vayanos, P.; Eric, R.; and Milind, T. 2018.
\newblock Designing Fair, Efficient, and Interpretable Policies for Prioritizing Homeless Youth for Housing Resources.
\newblock In van Hoeve, W.-J., ed., \emph{Integration of Constraint Programming, Artificial Intelligence, and Operations Research}, 35--51. Springer International Publishing.

\bibitem[{Bertsimas, Delarue, and Pauphilet(2024)}]{Bertsimas2024_DataImputation}
Bertsimas, D.; Delarue, A.; and Pauphilet, J. 2024.
\newblock {Simple Imputation Rules for Prediction with Missing Data: Theoretical Guarantees vs. Empirical Performance}.
\newblock \emph{Transactions on Machine Learning Research}.

\bibitem[{Brayne and Christin(2021)}]{Brayne2021}
Brayne, S.; and Christin, A. 2021.
\newblock Technologies of Crime Prediction: The Reception of Algorithms in Policing and Criminal Courts.
\newblock \emph{Social Problems}, 68: 608--624.

\bibitem[{Brown et~al.(2018)Brown, Cummings, Lyons, Carrion, and Watson}]{Brown2018}
Brown, M.; Cummings, C.; Lyons, J.; Carrion, A.; and Watson, D. 2018.
\newblock Reliability and validity of the Vulnerability Index-Service Prioritization Decision Assistance Tool (VI-SPDAT) in real-world implementation.
\newblock \emph{Journal of Social Distress and the Homeless}, 27.

\bibitem[{Chouldechova et~al.(2018)Chouldechova, Benavides-Prado, Fialko, and Vaithianathan}]{Chouldechova2018}
Chouldechova, A.; Benavides-Prado, D.; Fialko, O.; and Vaithianathan, R. 2018.
\newblock A case study of algorithm-assisted decision making in child maltreatment hotline screening decisions.
\newblock In Friedler, S.~A.; and Wilson, C., eds., \emph{Proceedings of the 1st Conference on Fairness, Accountability and Transparency}, volume~81 of \emph{Proceedings of Machine Learning Research}, 134--148. PMLR.

\bibitem[{{City of Hamilton}(2024)}]{Hamilton2024}
{City of Hamilton}. 2024.
\newblock {Encampment Protocol}.
\newblock https://www.hamilton.ca/sites/default/files/2024-08/Housing\_Encampment-Protocol-Revised-June2024.pdf.

\bibitem[{De~Jong(2024)}]{OrgCode_MovingForward}
De~Jong, I. 2024.
\newblock {A Message from OrgCode on the VI-SPDAT Moving Forward}.
\newblock \url{https://www.orgcode.com/blog/a-message-from-orgcode-on-the-vi-spdat-moving-forward}.

\bibitem[{Denton(2019)}]{Denton2019}
Denton, J. 2019.
\newblock {Will Algorithmic Tools Help or Harm the Homeless?}
\newblock \emph{Pacific Standard}.

\bibitem[{Eubanks(2019)}]{AutomatingInequality}
Eubanks, V. 2019.
\newblock \emph{{Automating Inequality: How High-Tech Tools Profile, Police, and Punish the Poor}}.
\newblock Picador.

\bibitem[{Flügge, Hildebrandt, and Møller(2021)}]{Flugge2021}
Flügge, A.~A.; Hildebrandt, T.; and Møller, N.~H. 2021.
\newblock Street-Level Algorithms and AI in Bureaucratic Decision-Making.
\newblock In \emph{Proceedings of the ACM on Human-Computer Interaction}, volume~5.

\bibitem[{Fried(2024)}]{Fried2024}
Fried, I. 2024.
\newblock {OpenAI says ChatGPT usage has doubled since last year}.
\newblock \emph{Axios}.

\bibitem[{{Homeless Individuals and Families Information System}(2024)}]{CanadaHIFIS2024}
{Homeless Individuals and Families Information System}. 2024.
\newblock {Coordinated Access in HIFIS}.
\newblock \url{https://homelessnesslearninghub.ca/wp-content/uploads/2024/05/HIFISCAGuide-EN-Final.pdf}.

\bibitem[{Jo et~al.(2023)Jo, Tang, Dullerud, Aghaei, Rice, and Vayanos}]{JoTang_Allocation_Incompatability}
Jo, N.; Tang, B.; Dullerud, K.; Aghaei, S.; Rice, E.; and Vayanos, P. 2023.
\newblock {Fairness in Contextual Resource Allocation Systems: Metrics and Incompatibility Results}.
\newblock In \emph{Proceedings of the Thirty-Seventh AAAI Conference on Artificial Intelligence}.

\bibitem[{Johnston, Blessenohl, and Vayanos(2020)}]{Johnston2020}
Johnston, C.~M.; Blessenohl, S.; and Vayanos, P. 2020.
\newblock {Preference Elicitation and Aggregation to Aid with Patient Triage during the COVID-19 Pandemic}.
\newblock \emph{Workshop on Participatory Approaches to Machine Learning}.

\bibitem[{Johnston et~al.(2023)Johnston, Vossler, Blessenohl, and Vayanos}]{Johnston2023}
Johnston, C.~M.; Vossler, P.; Blessenohl, S.; and Vayanos, P. 2023.
\newblock Deploying a Robust Active Preference Elicitation Algorithm on MTurk: Experiment Design, Interface, and Evaluation for COVID-19 Patient Prioritization.
\newblock In \emph{Proceedings of the 3rd ACM Conference on Equity and Access in Algorithms, Mechanisms, and Optimization}.

\bibitem[{Justin et~al.(2023)Justin, Aghaei, G{\'{o}}mez, and Vayanos}]{Justin2023}
Justin, N.; Aghaei, S.; G{\'{o}}mez, A.; and Vayanos, P. 2023.
\newblock {Learning Optimal Classification Trees Robust to Distribution Shifts}.
\newblock arXiv:2310.17772.

\bibitem[{Kallus and Zhou(2021)}]{Kallus2021}
Kallus, N.; and Zhou, A. 2021.
\newblock Minimax-Optimal Policy Learning Under Unobserved Confounding.
\newblock \emph{Management Science}, 67: 2870--2890.

\bibitem[{Karusala et~al.(2019)Karusala, Wilson, Vayanos, and Rice}]{Karusala2019}
Karusala, N.; Wilson, J.; Vayanos, P.; and Rice, E. 2019.
\newblock {The street-level realities of data practices in homeless services provision}.
\newblock In \emph{Proceedings of the ACM on Human-Computer Interaction}, volume 3, CSCW.

\bibitem[{Kern, Kim, and Zhou(2024)}]{Kern2024_MissingDataCATE}
Kern, C.; Kim, M.; and Zhou, A. 2024.
\newblock {Multi-Accurate CATE is Robust to Unknown Covariate Shifts}.
\newblock \emph{Transactions on Machine Learning Research}.

\bibitem[{KPMG(2024)}]{KPMG2024}
KPMG. 2024.
\newblock Current State Assessment Report: United Way of Greater Los Angeles Homeless Sector Recruitment Analysis.
\newblock Technical report, Homelessness Research Institute, National Alliance to End Homelessness.

\bibitem[{Kube, Das, and Fowler(2019)}]{Kube2019}
Kube, A.~R.; Das, S.; and Fowler, P.~J. 2019.
\newblock {Allocating Interventions Based on Predicted Outcomes: A Case Study on Homelessness Services}.
\newblock In \emph{Proceedings of the AAAI Conference on Artificial Intelligence}, volume~33, 622--629.

\bibitem[{Kube, Das, and Fowler(2023)}]{Kube2023}
Kube, A.~R.; Das, S.; and Fowler, P.~J. 2023.
\newblock {Fair and Efficient Allocation of Scarce Resources Based on Predicted Outcomes: Implications for Homeless Service Delivery}.
\newblock \emph{Journal of Artificial Intelligence Research}, 76(73): 1219--1245.

\bibitem[{Kumar and Yeoh(2023)}]{Kumar2023}
Kumar, A.; and Yeoh, W. 2023.
\newblock {Fairness in Scarce Societal Resource Allocation: A Case Study in Homelessness Applications}.
\newblock In \emph{Proceedings of the Workshop on Autonomous Agents for Social Good}.

\bibitem[{Kuo et~al.(2023)Kuo, Shen, Geum, Jones, Hong, Zhu, and Holstein}]{Kuo2023}
Kuo, T.-S.; Shen, H.; Geum, J.; Jones, N.; Hong, J.~I.; Zhu, H.; and Holstein, K. 2023.
\newblock {Understanding Frontline Workers' and Unhoused Individuals' Perspectives on AI Used in Homeless Services}.
\newblock In \emph{Proceedings of the 2023 CHI Conference on Human Factors in Computing Systems}.

\bibitem[{Lee et~al.(2019)Lee, Kusbit, Kahng, Kim, Yuan, Chan, See, Noothigattu, Lee, Psomas, and Procaccia}]{Lee2019}
Lee, M.~K.; Kusbit, D.; Kahng, A.; Kim, J.~T.; Yuan, X.; Chan, A.; See, D.; Noothigattu, R.; Lee, S.; Psomas, A.; and Procaccia, A.~D. 2019.
\newblock {WeBuildAI: Participatory framework for algorithmic governance}.
\newblock In \emph{Proceedings of the ACM on Human-Computer Interaction}, volume~3, 1--35.

\bibitem[{{Los Angeles Homeless Services Authority}(2017)}]{LAHSA_Women}
{Los Angeles Homeless Services Authority}. 2017.
\newblock {Report and Recommendations of the Ad Hoc Committee on Women and Homelessness}.
\newblock \url{https://cao.lacity.gov/Homeless/hsc20170928c.pdf}.

\bibitem[{{Los Angeles Homeless Services Authority}(2018)}]{LAHSA_BlackPeople}
{Los Angeles Homeless Services Authority}. 2018.
\newblock {Report and Recommendations of the Ad Hoc Committee on Black People Experiencing Homelessness}.
\newblock \url{https://www.lahsa.org/documents?id=2823-report-and-recommendations-of-the-ad-hoc-committee-on-black-people-experiencing-homelessness.pdf}.

\bibitem[{{Los Angeles Homeless Services Authority}(2024{\natexlab{a}})}]{LAHSA_HomelessCount2024}
{Los Angeles Homeless Services Authority}. 2024{\natexlab{a}}.
\newblock 2024 Greater Los Angeles Homeless Count Results (Long Version).
\newblock \url{https://www.lahsa.org/documents?id=8164-2024-greater-los-angeles-homeless-count-results-long-version-.pdf}.

\bibitem[{{Los Angeles Homeless Services Authority}(2024{\natexlab{b}})}]{LAHSA_HousingInventory2024}
{Los Angeles Homeless Services Authority}. 2024{\natexlab{b}}.
\newblock 2024 Housing Inventory Count.
\newblock \url{https://www.lahsa.org/documents?id=8162-2024-housing-inventory-count.xlsx}.

\bibitem[{{Los Angeles Homeless Services Authority}(2024{\natexlab{c}})}]{LAHSA_matching}
{Los Angeles Homeless Services Authority}. 2024{\natexlab{c}}.
\newblock {Coordinated Entry System Assessment/Matching Guidance {\&} How to Administer the CES Assessment (VI-SPDAT)}.
\newblock \url{https://www.lahsa.org/documents?id=7969-ces-guidance-and-vi-spdat-training.pdf}.

\bibitem[{{Los Angeles Homeless Services Authority}(2024{\natexlab{d}})}]{LAHSA_centralized_matching}
{Los Angeles Homeless Services Authority}. 2024{\natexlab{d}}.
\newblock {Webinar Centralized Matching Presentation}.
\newblock \url{https://www.lahsa.org/documents?id=8235-webinar-centralized-matching-presentation.pdf}.

\bibitem[{Mashiat et~al.(2022)Mashiat, Gitiaux, Rangwala, Fowler, and Das}]{Mashiat2022}
Mashiat, T.; Gitiaux, X.; Rangwala, H.; Fowler, P.; and Das, S. 2022.
\newblock {Trade-offs between Group Fairness Metrics in Societal Resource Allocation}.
\newblock In \emph{Proceedings of the 2022 ACM Conference on Fairness, Accountability, and Transparency}, 1095--1105.

\bibitem[{Moon and Guha(2024)}]{Moon2024}
Moon, E. S.~Y.; and Guha, S. 2024.
\newblock {A Human-Centered Review of Algorithms in Homelessness Research}.
\newblock In \emph{Proceedings of the 2024 CHI Conference on Human Factors in Computing Systems}.

\bibitem[{Moon et~al.(2025)Moon, Saxena, Das, and Guha}]{Moon2025}
Moon, E. S.-Y.; Saxena, D.; Das, D.; and Guha, S. 2025.
\newblock The Datafication of Care in Public Homelessness Services.
\newblock In \emph{Proceedings of the 2025 CHI Conference on Human Factors in Computing Systems}, 1--16.

\bibitem[{Moses(2023)}]{Moses2023}
Moses, J. 2023.
\newblock Working in Homeless Services: A Survey of the Field.
\newblock Technical report, Homelessness Research Institute, National Alliance to End Homelessness.

\bibitem[{Møller, Shklovski, and Hildebrandt(2020)}]{Moller2020}
Møller, N.~H.; Shklovski, I.; and Hildebrandt, T.~T. 2020.
\newblock Shifting Concepts of Value: Designing Algorithmic Decision-Support Systems for Public Services.
\newblock In \emph{Proceedings of the 11th Nordic Conference on Human-Computer Interaction: Shaping Experiences, Shaping Society}.

\bibitem[{{Orgcode}(2014)}]{Orgcode2014}
{Orgcode}. 2014.
\newblock {Vulnerability Index (VI) and Service Prioritization Decision Assistance Tool (SPDAT)}.

\bibitem[{Padgett(2008)}]{Padgett2008}
Padgett, D. 2008.
\newblock \emph{Qualitative methods in social work research}.
\newblock Sage Publications, 2nd edition.

\bibitem[{Rahmattalabi et~al.(2022)Rahmattalabi, Vayanos, Dullerud, and Rice}]{Rahmattalabi2022}
Rahmattalabi, A.; Vayanos, P.; Dullerud, K.; and Rice, E. 2022.
\newblock {Learning Resource Allocation Policies from Observational Data with an Application to Homeless Services Delivery}.
\newblock In \emph{Proceedings of the 2022 ACM Conference on Fairness, Accountability, and Transparency}, 1240--1256.

\bibitem[{Reamer(2023)}]{Reamer2023}
Reamer, F.~G. 2023.
\newblock {Artificial Intelligence in Social Work: Emerging Ethical Issues}.
\newblock \emph{International Journal of Social Work Values and Ethics}, 20.

\bibitem[{Rice et~al.(2023)Rice, Milburn, Vayanos, Rountree, Hill, Petering, Blackwell, Santillano, Onasch-Vera, Nadel, Tang, Aghaei, Hsu, and Petry}]{Rice2023CESTTRR}
Rice, E.; Milburn, N.; Vayanos, P.; Rountree, J.; Hill, C.; Petering, R.; Blackwell, B.; Santillano, R.; Onasch-Vera, L.; Nadel, H.~W.; Tang, B.; Aghaei, S.; Hsu, H.; and Petry, L. 2023.
\newblock CESTTRR: Coordinated Entry System Triage Tool Research and Refinement.
\newblock Technical report.

\bibitem[{Saxena et~al.(2021)Saxena, Badillo-Urquiola, Wisniewski, and Guha}]{Saxena2021}
Saxena, D.; Badillo-Urquiola, K.; Wisniewski, P.~J.; and Guha, S. 2021.
\newblock A Framework of High-Stakes Algorithmic Decision-Making for the Public Sector Developed through a Case Study of Child-Welfare.
\newblock In \emph{Proceedings of the ACM on Human-Computer Interaction}, volume~5.

\bibitem[{Saxena and Guha(2024)}]{Saxena2024}
Saxena, D.; and Guha, S. 2024.
\newblock Algorithmic Harms in Child Welfare: Uncertainties in Practice, Organization, and Street-level Decision-making.
\newblock \emph{ACM Journal on Responsible Computing}, 1: 1--32.

\bibitem[{Showkat et~al.(2023)Showkat, Smith, Lingqing, and To}]{Showkat2023}
Showkat, D.; Smith, A.~D.; Lingqing, W.; and To, A. 2023.
\newblock {``Who is the right homeless client?": Values in Algorithmic Homelessness Service Provision and Machine Learning Research}.
\newblock In \emph{Proceedings of the 2023 CHI Conference on Human Factors in Computing Systems}.

\bibitem[{Stapleton et~al.(2022)Stapleton, Lee, Qing, Wright, Chouldechova, Holstein, Wu, and Zhu}]{Stapleton2022ImaginingStakeholders}
Stapleton, L.; Lee, M.~H.; Qing, D.; Wright, M.; Chouldechova, A.; Holstein, K.; Wu, Z.~S.; and Zhu, H. 2022.
\newblock {Imagining new futures beyond predictive systems in child welfare: A qualitative study with impacted stakeholders}.
\newblock In \emph{Proceedings of the 2022 ACM Conference on Fairness, Accountability, and Transparency}, 1162--1177.

\bibitem[{Tang et~al.(2023)Tang, Ko{\c{c}}yi{\u{g}}it, Rice, and Vayanos}]{Tang2023}
Tang, B.; Ko{\c{c}}yi{\u{g}}it, {\c{C}}.; Rice, E.; and Vayanos, P. 2023.
\newblock {Learning Optimal and Fair Policies for Online Allocation of Scarce Societal Resources from Data Collected in Deployment}.
\newblock arXiv:2311.13765.

\bibitem[{Tang et~al.(2024)Tang, Zhi, Kuo, Kainaroi, Northup, Holstein, Zhu, Heidari, and Shen}]{Tang2024}
Tang, N.; Zhi, J.; Kuo, T.~S.; Kainaroi, C.; Northup, J.~J.; Holstein, K.; Zhu, H.; Heidari, H.; and Shen, H. 2024.
\newblock {AI Failure Cards: Understanding and Supporting Grassroots Efforts to Mitigate AI Failures in Homeless Services}.
\newblock In \emph{Proceedings of the 2024 ACM Conference on Fairness, Accountability, and Transparency}, 713--732.

\bibitem[{{The County of Santa Clara}(2024)}]{SantaClaraCES}
{The County of Santa Clara}. 2024.
\newblock {Coordinated Entry}.
\newblock \url{https://scc.bitfocus.com/coordinated-entry}.

\bibitem[{Toros and Flaming(2018)}]{Toros2018}
Toros, H.; and Flaming, D. 2018.
\newblock {Prioritizing Homeless Assistance Using Predictive Algorithms: An Evidence-Based Approach}.
\newblock \emph{Cityscape}, 20(1).

\bibitem[{Tracey and Garcia(2024{\natexlab{a}})}]{Tracey2024a}
Tracey, P.; and Garcia, P. 2024{\natexlab{a}}.
\newblock After automation: Homelessness prioritization algorithms and the future of care labor.
\newblock \emph{Big Data and Society}, 11.

\bibitem[{Tracey and Garcia(2024{\natexlab{b}})}]{Tracey2024}
Tracey, P.; and Garcia, P. 2024{\natexlab{b}}.
\newblock {Intermediation: Algorithmic Prioritization in Practice in Homeless Services}.
\newblock In \emph{Proceedings of the ACM on Human-Computer Interaction}, volume 8, CSCW2.

\bibitem[{{U.S. Department of Housing and Urban Development}(2017)}]{HUD2017}
{U.S. Department of Housing and Urban Development}. 2017.
\newblock CPD-17-01 Notice Establishing Additional Requirements for a Continuum of Care Centralized or Coordinated Assessment System.
\newblock Technical report.

\bibitem[{Vayanos et~al.(2020)Vayanos, Ye, McElfresh, Dickerson, and Rice}]{Vayanos2020}
Vayanos, P.; Ye, Y.; McElfresh, D.~C.; Dickerson, J.~P.; and Rice, E. 2020.
\newblock {Robust Active Preference Elicitation}.
\newblock arXiv:2003.01899.

\bibitem[{Watson-Daniels et~al.(2023)Watson-Daniels, Barocas, Hofman, and Chouldechova}]{Watson-Daniels2023}
Watson-Daniels, J.; Barocas, S.; Hofman, J.~M.; and Chouldechova, A. 2023.
\newblock {Multi-Target Multiplicity: Flexibility and Fairness in Target Specification under Resource Constraints}.
\newblock In \emph{Proceedings of the 2023 ACM Conference on Fairness, Accountability, and Transparency}, 297--311.

\end{thebibliography}

%
\appendix 

\section{Interview Protocol}\label{sec:InterviewProtocol}
\begin{enumerate}
    \item [(Q1)] What is your age group? Gender? Race/ethnicity?
    \item [(Q2)] Can you give me a brief overview of your advisory role [duties as a SPA lead]?
    \item [(Q3)] How long have you been at your current organization? How long have you been working in the homelessness field?
    \item [(Q4)] Can you tell me how a client’s VI-SPDAT score is used during case conferencing to match them to a housing resource?
    \item  [(Q5)]What other information is used to determine who gets a particular housing resource (e.g., HMIS data)? Is there information used outside of what is stored in the HMIS? If so, what is it?
    \item [(Q6)] Recently, it seems like public awareness and use of AI has exploded with the introduction of chatbots like ChatGPT. However, there are lots of AI tools used for decision-making that are already deeply integrated into our daily lives. For example, AI is used to recommend shows to watch on Netflix or products to buy on Amazon, to find the fastest way to travel on Google Maps, or to follow the rules of the road for self-driving cars. Do you think about AI much in your daily life? In what sense? Have you heard about AI much in the news (either in a positive or negative light)? What was the topic?
    \item [(Q7)] Have you ever thought about how AI can use HMIS data to make improved housing decisions during case conferencing? If so, in what specific ways? (\textit{If they don’t have an answer)} For example, AI can be used to determine which clients can stabilize their housing situation by receiving a TLS vs. receiving PSH. Does this idea sound appealing to you? Or do you have hesitations? Why?
    \item [(Q8)] Do you think using AI for making housing decisions could lead to mostly positive change, mostly negative change, or a mix? Why?

[SHOW INTRODUCTORY SLIDES \& SLIDES ON CLIENT OUTCOMES]

    \item [(Q9)] What other potential client outcomes (either quantitative or qualitative) are important to you?

\item [(Q10)] Now let’s look at an example of how choosing a measurable outcome can have unintended consequences related to fairness. Suppose we focus on getting people off the street who are most likely to die there. Resources would then probably go to clients who have serious medical conditions. However, studies have shown that non-White individuals are less likely to be diagnosed for various conditions compared to White individuals. 
This most likely means that more White clients will get housing resources, 	even though non-White clients may need the support just as much. 
Does this example make sense to you? Do you see that there’s a relationship between defining client outcomes and fairness? Is there anything from this example that concerns you? What about this relationship between outcomes and fairness is unexpected and why? Do any similar examples come to mind that you’re aware of in the current allocation system?

[SHOW SLIDES ON FAIRNESS]

\item [(Q11)] What about these results surprises you and why? What about this confirms your previous understanding of fairness (either in the ways we discussed or in other ways) in the current system and why?

\item [(Q12)]What else is considered “fair” to you that you would want to see in a policy?

[SHOW SLIDES ON TRANSPARENCY]

\item [(Q13)]What else comes to mind in terms of transparency of the current system (either in terms of the scoring system or housing resource allocation as a whole)? What aspects are transparent or that could be more transparent?

\item [(Q14)] What about these results is surprising and why? What about this confirms your previous understanding of transparency in the current system and why?

\item [(Q15)] Now that we’ve discussed more examples concerning transparency, is there anything you would want to see or know from a policy in terms of this?

[SHOW SLIDES ON PROPOSED AI SYSTEM]

\item [(Q16)] Is there anything about these results that is unexpected and why? What about it confirms your previous understanding of how AI could be used for housing and why?

\item [(Q17)] For each of the following proposed features of an AI system, is this information you’d like to know? In what ways is it an improvement or not an improvement over the current allocation system? Why?
\begin{itemize}
    \item Clients are matched in order of arrival	
    \item Clients know their place in line
    \item Clients with the same experiences are routed to the same resources (with some leeway)
    \item Clients know estimated wait times
\end{itemize}

\item [(Q18)] After having this discussion, what is your greatest concern about using AI to allocate housing resources to clients? Why? What do you think could be the greatest value or benefit in using AI to allocate housing resources to clients? Why?

\item [(Q19)] Any final thoughts or anything we didn’t cover on using AI to build such an allocation system?

\item [(Q20)] After our discussion, do you think using AI for making housing decisions could lead to mostly positive change, mostly negative change, or a mix? Why?
\end{enumerate}

\section{Codes}
The following are the codes generated from our thematic analysis:
\begin{itemize}
    \item Perceptions of AI
    \item Complexity of the Housing System
    \item Data Issues
    \item Decision-making Process
    \item Fairness
    \item Other AI Solutions
\end{itemize}

\section{Information Sheet}

My name is Caroline Johnston and I am a Ph.D. student at the University of Southern California. I am part of the research team for the CES Triage Tool Research \& Refinement (CESTTRR) Project.

As part of my dissertation, I am conducting a research study to identify the current challenges with the CES tools and to gauge stakeholders' openness to data-driven (AI) approaches for housing allocation tools. The name of this research study is “Eliciting Policymakers' Preferences for Designing Data-driven Approaches for Allocating Scarce Resources to Individuals Experiencing Homelessness.” I am seeking your participation in this study.

Your participation is completely voluntary, and I will address your questions or concerns at any point before or during the study. 

You may be eligible to participate in this study if you meet the following criteria:
\begin{enumerate}
    \item You are over 18 years old.
    \item You are English-speaking.
    \item You are a member of [REDACTED FOR CONFIDENTIALITY] or [REDACTED FOR CONFIDENTIALITY].
\end{enumerate}
If you decide to participate in this study, you will be asked to do the following activities:
\begin{enumerate}
    \item Participate in a 1:1 60-minute interview over a video conferencing program (e.g., Skype, Zoom, Google Hangouts).
\end{enumerate}
After you complete the interview, if you are allowed to receive payment in this capacity, you will receive a \$100 Amazon gift card via email.

I will publish the results in my thesis and/or a publication. Participants will not be identified in the results. I will take reasonable measures to protect the security of all your personal information. De-identified transcripts and separate, coded data with personal identifying information will be kept on a secure server housed at the University of Southern California and retained for future use. I may share your data, de-identified with other researchers in the future. You may be contacted by a member of the research team in the future to take part in follow-up studies.

If you have any questions about this study, please contact me: caroline.johnston@usc.edu. If you have any questions about your rights as a research participant, please contact the University of Southern California Institutional Review Board at (323) 442-0114 or email hrpp@usc.edu.
\newpage
\section{Interview Slides Material}
\includegraphics[scale=.7]{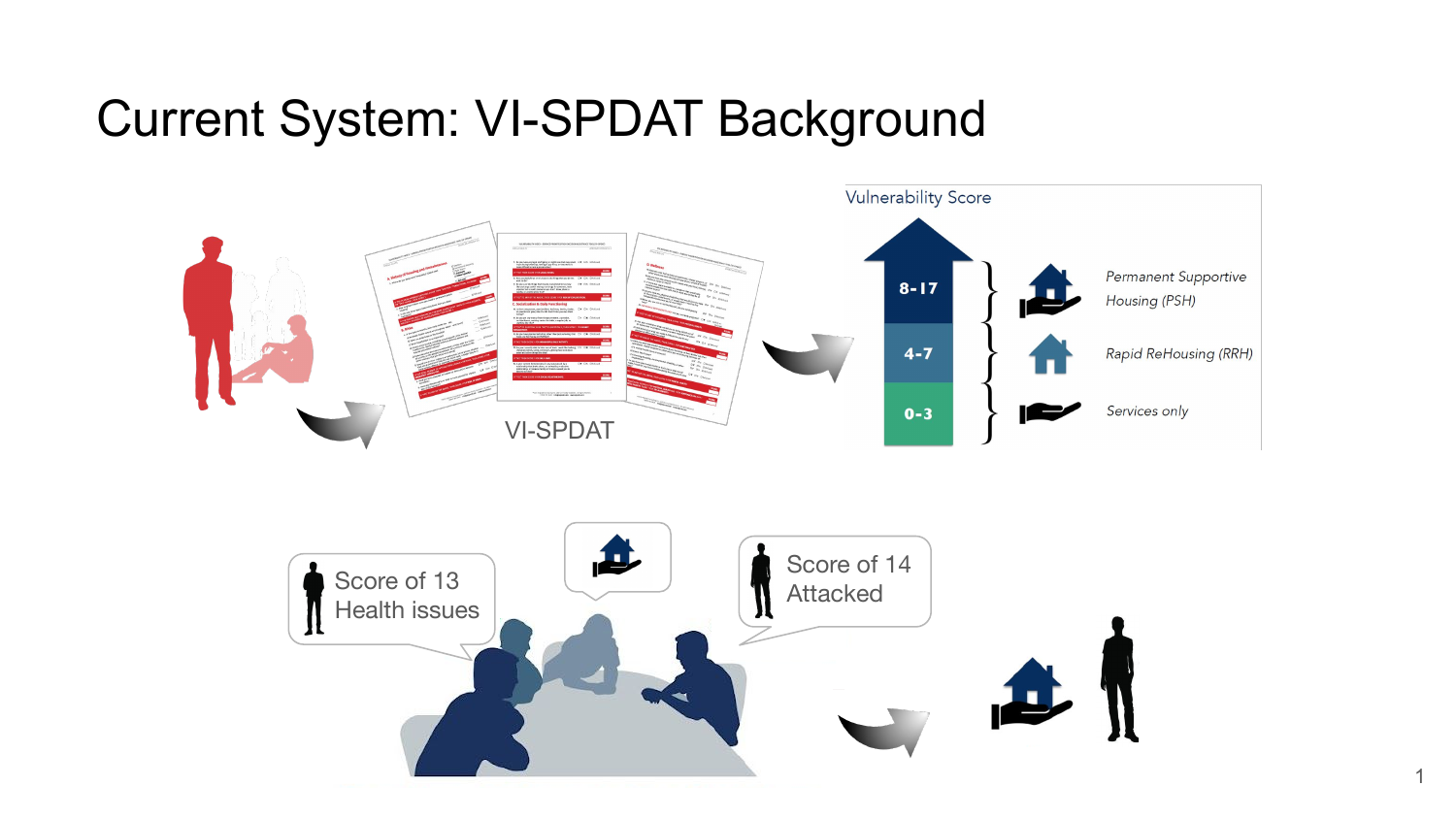}
\includegraphics[scale=.7]{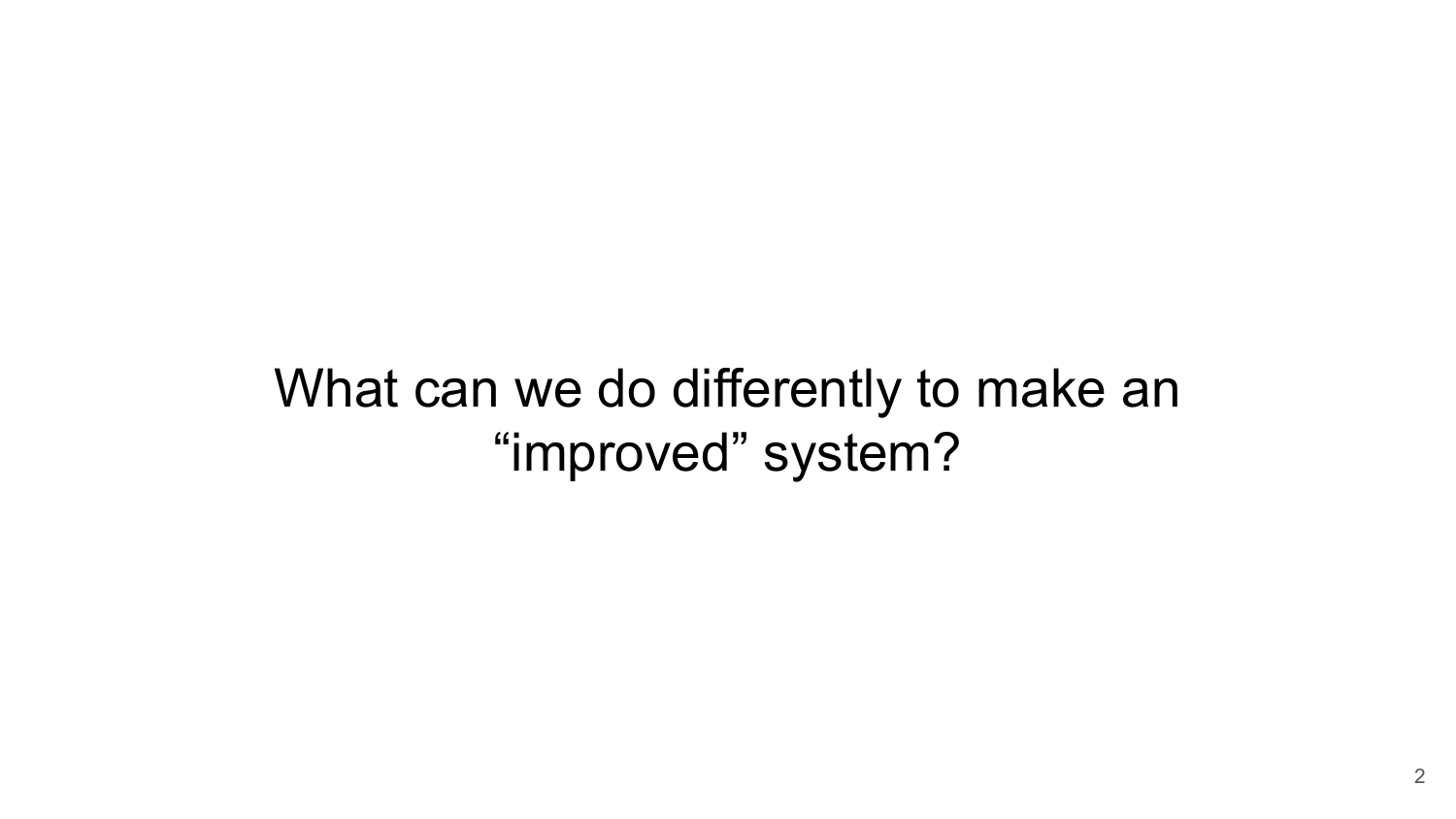}
\clearpage


\section*{Supplementary material}

**Interview Slides Material**

# Current System: VI-SPDAT Background

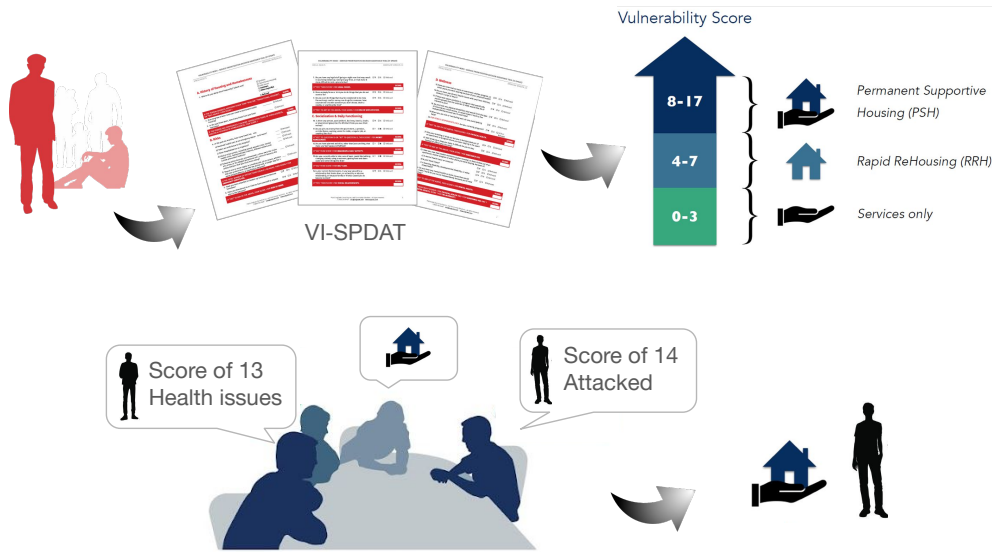

1

# What can we do differently to make an "improved" system?

2

## Designing Policies to Improve Clients' Measurable Outcomes

- Extremely scarce housing resources
    - Who gets the resources?
- Possible measurable client outcomes:

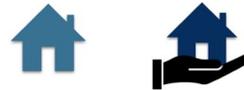

| Minimize number of days clients spend on the street | Get most clients off the streets |
|---|---|
| Get clients off the street who are most likely to die there | Minimize clients' emergency room visits or jail time |

3

## Designing Policies that are "Fair"

**Allocation Parity**

- Clients **receive resources** at equal rates across protected groups

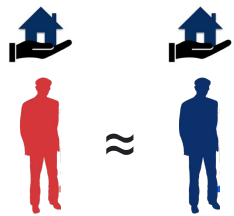

**Outcome Parity**

- Clients **exit homelessness** (positive outcome) at equal rates across protected groups

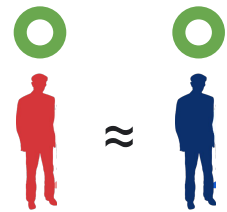

4

## Designing Policies that are "Fair"

**Allocation Parity**

- Clients **receive resources** at equal rates across protected groups

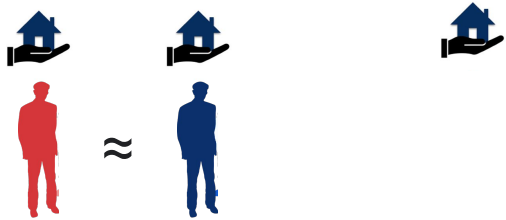

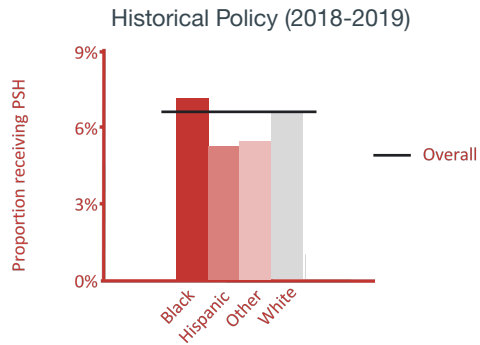

Observable disparities across racial groups in **allocation**

5

## Designing Policies that are "Fair"

**Outcome Parity**

- Clients **exit homelessness** (positive outcome) at equal rates across protected groups

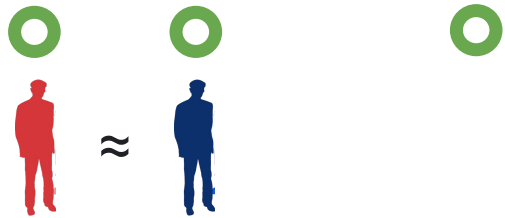

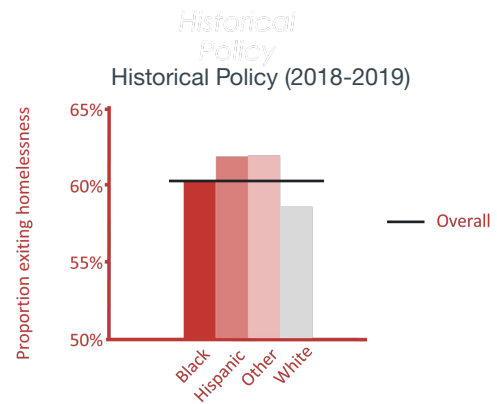

Observable disparities across racial groups in **outcomes**

6

# Current System: Transparency

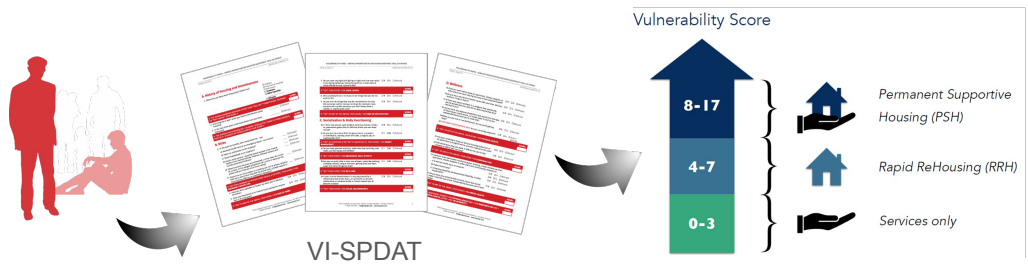

# Current System: Transparency

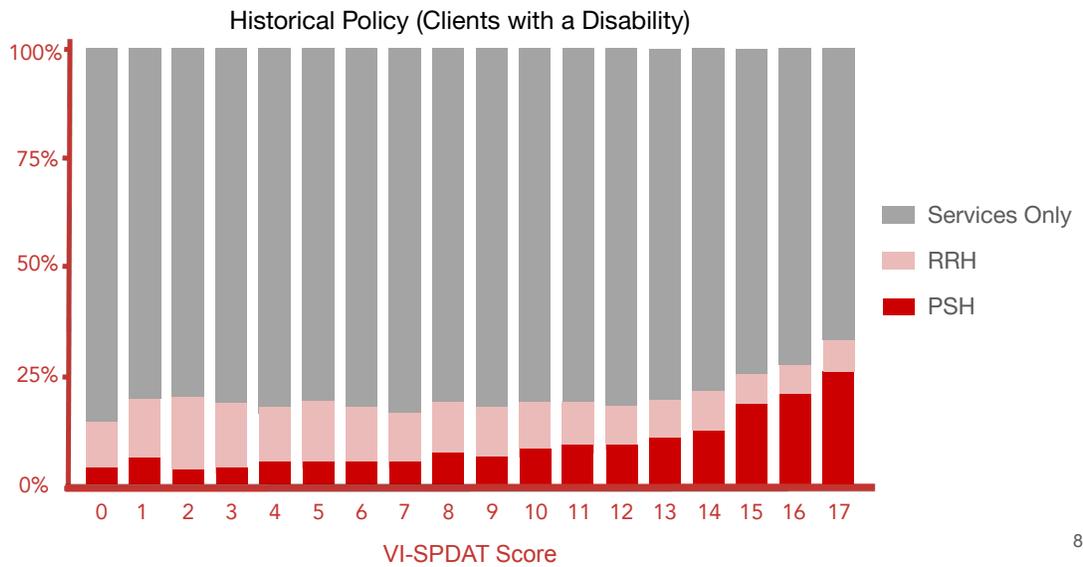

## Current System: Transparency

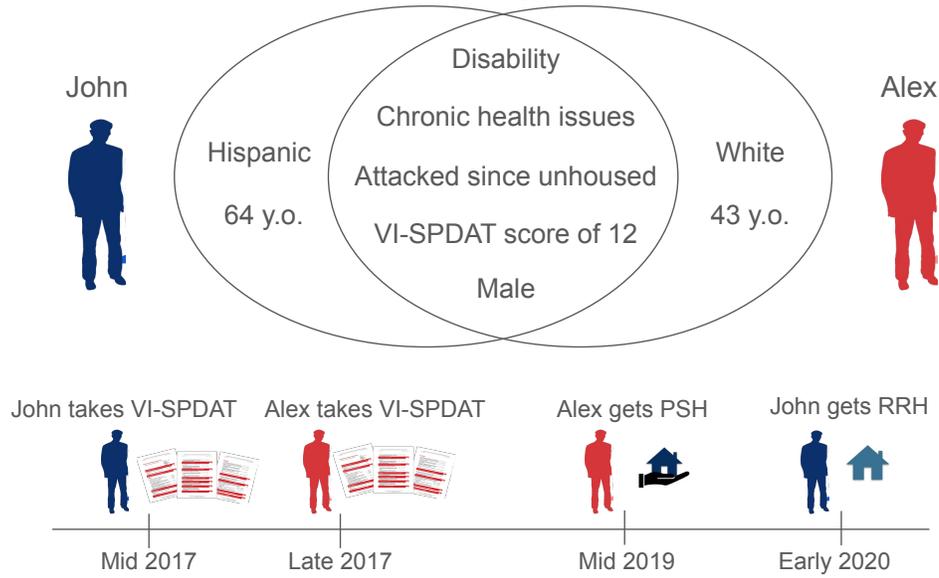

What can AI do for the resource allocation process?

## AI-driven Allocation Policy: Outcomes & Fairness

- Assign resources **optimally** using HMIS data to **maximize** the **number of clients** that **successfully exit homelessness**

- Guarantee either **outcome parity** or **allocation parity**

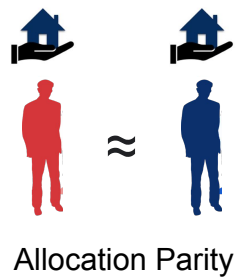
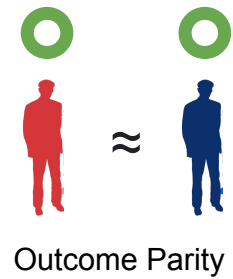

Allocation Parity          Outcome Parity

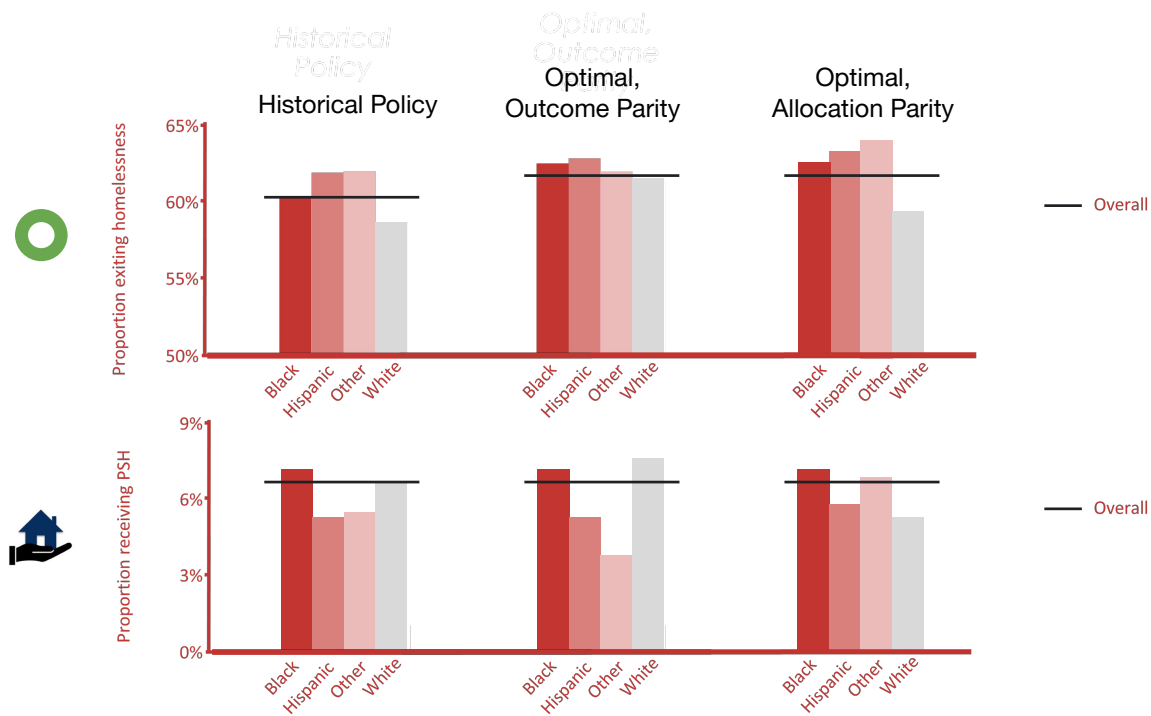

11

12

## AI-driven Allocation Policy: Transparency

- Policy assigns individuals with the **same experiences** to a queue for the **same type of resource**, **in order of arrival**

- Clients will **know their place in line** & **estimated wait times**

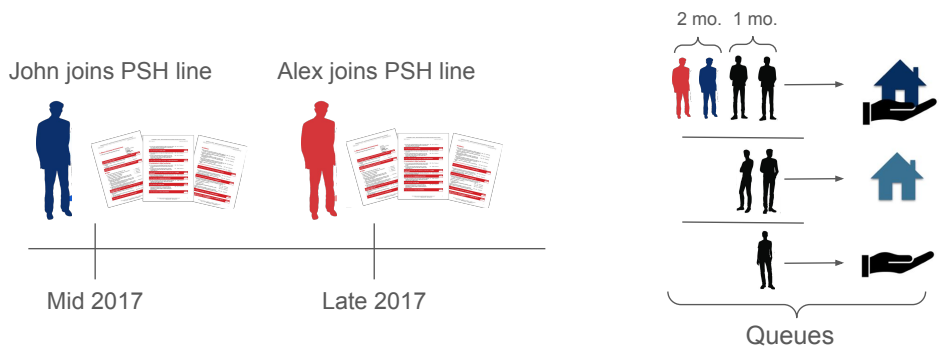

13



\end{document}